\documentclass[superscriptaddress,prd,showpacs]{revtex4}
\usepackage{amssymb,amsmath,graphicx,amscd,xcolor}

\begin{document}

\title{A Model for Lightcone Fluctuations due to Stress Tensor Fluctuations}

\author{C. H. G. \surname{Bessa}}
\email{carlos@cosmos.phy.tufts.edu}
\affiliation{Institute of Cosmology, Department of Physics and Astronomy, \\
Tufts University, Medford, Massachusetts 02155, USA}
\affiliation{Departamento de F\'{\i}sica, Universidade Federal da Para\'{\i}ba, \\
Caixa Postal 5008, CEP 58051-970, Jo\~ao Pessoa, PB, Brazil}
\author{V. A. \surname{De Lorenci}}
\email{delorenci@unifei.edu.br}
\affiliation{Instituto de F\'{\i}sica e Qu\'{\i}mica,    Universidade Federal de Itajub\'a, \\
Itajub\'a, Minas Gerais 37500-903, Brazil}
\author{L. H. \surname{Ford}}
\email{ford@cosmos.phy.tufts.edu}
\affiliation{Institute of Cosmology, Department of Physics and Astronomy, \\
Tufts University, Medford, Massachusetts 02155, USA}
\author{C. C. H. \surname{Ribeiro}}
\email{caiocesarribeiro@unifei.edu.br}
\affiliation{Instituto de F\'{\i}sica e Qu\'{\i}mica,    Universidade Federal de Itajub\'a, \\
Itajub\'a, Minas Gerais 37500-903, Brazil}

\begin{abstract}
We study a model for quantum lightcone fluctuations in which vacuum fluctuations of the electric field
and of the squared electric field in a nonlinear dielectric material produce variations in the flight times of
probe pulses. When this material has a non-zero third order polarizability, the flight time variations
arise from squared electric field fluctuations, and are analogous to effects expected when the stress
tensor of a quantized field drives passive spacetime geometry fluctuations. We also discuss the dependence
of the squared electric field fluctuations upon the geometry of the material, which in turn determines a
sampling function for averaging the squared electric field along the path of the pulse. This allows us to estimate
the probability of especially large fluctuations, which is a measure of the probability distribution for
quantum stress tensor fluctuations.
\end{abstract}

		
\maketitle
\baselineskip=14pt	

\section{introduction}

Light propagation in a nonlinear dielectric may be used to model various subtle effects involving quantum
theory and gravity. These include lightcone fluctuations~\cite{FDMS13,BDFS15} and the effects of quantum 
stress tensor expectation values in semiclassical gravity~\cite{BDF14}.  A fluctuating electric field in a
nonlinear material causes fluctuations of the effective speed of light of probe pulses, and is analogous to the effects
of spacetime geometry fluctuations on light propagation. This analogy was developed in Ref.~\cite{FDMS13},
where the source of the fluctuations was a squeezed state of the electromagnetic field, and in Ref.~\cite{BDFS15},
where the effects of vacuum fluctuations of the electric field were investigated. In both cases, linear fluctuations
of the electric field were treated, which models the lightcone fluctuations produced by active gravitational field
fluctuations. These are the fluctuations of the dynamical degrees of freedom of gravity itself, as opposed to
the passive fluctuations of gravity, driven by quantum stress tensor fluctuations. One of the purposes of the present 
paper will be to develop a model for passive spacetime geometry fluctuations. This will involve a study of the
fluctuations of the time averaged squared electric field, which is of interest in its own right. 

A second purpose of this paper will be a further study of switched fluctuations of quantum fields. The vacuum 
fluctuations of quantum field operators are only meaningful if the operators have been averaged in time or 
in spacetime with a smooth sampling function.  In the case of linear fields, such as the electric field, the
associated probability distribution is Gaussian. Some effects of the time averaged electric field were discussed
in Refs.~\cite{BDFS15,HF15}. In the latter paper, it was shown that simple arguments may be used to estimate
the one loop QED corrections to potential scattering by electrons. The fluctuations of quadratic field operators,
such as the stress tensor or the squared electric field, are more subtle, and are associated with non-Gaussian
probability distributions~\cite{FFR10,FFR12,FF15}. These distributions typically fall more slowly than a
Gaussian function, increasing the probability of large fluctuations, and depend sensitively upon the choice
of sampling function. It was argued in  Ref.~\cite{BDFS15} that the sampling function for vacuum fluctuations
in a dielectric can depend upon the geometry of the material. This idea will be further developed here, where
we will consider a broader class of functions than the Lorentzian function used in Ref.~\cite{BDFS15}.

The outline of this paper is as follows: Section~\ref{sec:flight} will first briefly review classical light propagation
in a nonlinear material, and then address the effects of switched vacuum fluctuations of the electric field and the
squared electric field. A convenient choice of switching function will be introduced in Sec.~\ref{sec:switch}.
Some numerical estimates of the magnitude of the flight time fluctuations will be given in  Sec.~\ref{sec:estimates}.
The probability distribution for the flight time fluctuations will be discussed in Sec.~\ref{sec:prob}. Our results 
will be summarized and discussed Sec.~\ref{sec:sum}.
Throughout this paper we use Lorentz-Heaviside units with $c=\hbar=1$.

\section{Flight time in a nonlinear optical material}
\label{sec:flight}
\subsection{Classical propagation speed}

A nonlinear dielectric material is one where the electric polarization vector is a nonlinear function of the electric field,
and can be written as~\cite{boyd2008}
\begin{equation}
P_i = \chi_{ij}^{(1)} E_{j} + \chi_{ijk}^{(2)} E_{j}E_{k} + \chi_{ijkl}^{(3)} E_{j}E_{k}E_{l} + \cdots  \,.
\label{eq:pol}
\end{equation}
Here repeated indices are summed upon, and  $\chi_{ij}^{(1)}$,  $\chi_{ijk}^{(2)}$, and $\chi_{ijkl}^{(3)}$
are the first, second, and third order susceptibility tensors, respectively. The second and higher order susceptibilities lead to 
a nonlinear wave equation for the electric field. 
We wish to investigate the flight time of a probe pulse propagating through a slab of optical material when second and third 
order coefficients of the susceptibility tensor are included. These nonlinearities of the medium couple to an external applied 
electric field $E^0_i(\textbf{x},t)$, here called the background field. The electric field associated with the probe pulse is 
denoted by the vector $\textbf{E}^1$, which we choose to be polarized in the $z$-direction and propagating in the $x$-direction, 
i.e., $\textbf{E}^1=E^1(x,t){\bf \hat{z}}$. Furthermore, we assume that the probe field is smaller in magnitude than the
background field, but more rapidly varying. That is, 
\begin{equation}
|E^1| \ll |E^0| \,,
\label{eq:C1}
\end{equation}
 but 
 \begin{equation}
 |\nabla E^0/E^0| \ll |\nabla E^1/E^1|\, .
\label{eq:C2}
\end{equation}
In this case,  $E^1$ obeys a linearized wave equation~\cite{FDMS13},
\begin{equation}
\frac{\partial^2 E^1}{\partial x^2} - \frac{1}{v_{ph}^2} \frac{\partial^2 E^{1}}{\partial t^2}  = 0\,.
\label{eq:we1}
\end{equation}
Here $v_{ph}$ is the phase velocity of the wave, which is given by
\begin{equation}
v_{ph}^2=\frac{1}{{n_p}^2}\left[1+2\gamma_i E_i^{0}+3\gamma_{ij} E_i^{0}E_j^{0}\right]^{-1},
\label{eq:vph}
\end{equation}
where $n_p= \bigl(1+\chi^{(1)}_{zz}\bigr)^{1/2}$ is the refractive index of the medium measured by the probe pulse when only 
linear effects take place, and we define the coefficients
\begin{align}
\gamma_i& = \frac{1}{n_p^2}\left(\frac{\chi^{(2)}_{zzi}+\chi^{(2)}_{ziz}}{2}\right),\\
\gamma_{ij}& = \frac{1}{n_p^2}\left(\frac{\chi^{(3)}_{zzij}+\chi^{(3)}_{zizj}+\chi^{(3)}_{zijz}}{3}\right).
\label{e2}
\end{align}
Equation~(\ref{eq:vph}) shows that the background field couples to the nonlinearities of the medium, affecting the  
velocity of the waves propagating through it. 

We will assume that dispersion can be ignored, so that the group velocity of a wavepacket is approximately equal 
to the phase velocity. In this case, the flight time of a probe pulse  traveling a distance $d$ in the x-direction will be 
given by
\begin{align}
t_d = n_p\int_0^d\left[1+\gamma_iE_i^0(\textbf{x},t)+\mu_{ij} E^0_i(\textbf{x},t)E^0_j(\textbf{x},t)\right]\,dx,
\label{td}
\end{align} 
with
\begin{align}
\mu_{ij} = \frac{1}{2}\left(3\gamma_{(ij)}-\gamma_i\gamma_j\right).
\label{mu}
\end{align}
Here the parenthesis enclosing two indices denotes symmetrization, i.e., $2\gamma_{(ij)}=\gamma_{ij}+\gamma_{ji}$.
In writing Eq.~(\ref{td}), we have assumed that the nonlinear effects are small, so that we may Taylor expand 
$1/v_{ph}$ from Eq.~(\ref{eq:vph}) to first order in $\gamma_{(ij)}$ and second order in $\gamma_{i}$. 
In addition, we take the integrand in Eq.~(\ref{td}) to be evaluated at   $t = n_p \, x$, which is the worldline of a pulse
traveling at speed $1/n_p$, that determined by the linear susceptibility.

\subsection{Vacuum fluctuations and switching}

In this paper, we will follow Ref.~\cite{BDFS15} and study the effects of vacuum electric field fluctuations as the
background field. In this case, $\textbf{E}^0$ becomes the quantized electric field operator, and $t_d$ defined in
Eq.~(\ref{td}) becomes an operator, where the term quadratic in $\textbf{E}^0$  is understood to be normal ordered,
$ E^0_i(\textbf{x},t)E^0_j(\textbf{x},t) \rightarrow : E^0_i(\textbf{x},t)E^0_j(\textbf{x},t):$. This leads to a finite mean
flight time, which in the vacuum state is, to leading order,
\begin{equation}
 \langle t_d\rangle = n_p\, d\,.
 \label{eq:mean}
 \end{equation}
 Our primary interest in this paper will be in the variance of the flight time,
 \begin{equation}
 (\Delta t_d)^2 = \langle t_d{}^2\rangle - \langle t_d\rangle^2\,.
 \label{eq:var}
 \end{equation}
 Note that this quantity is independent of the choice of vacuum state with respect to which normal ordering is
 performed. A change in the state has the effect of adding a c-number, $C$, to the operator $t_d$, so that
 $t_d \rightarrow t_d + C$. It is easily verified that the right-hand-side of Eq.~(\ref{eq:var}) is unchanged. A
 change of vacuum state can slightly change the mean time delay, $\langle t_d \rangle$, but  does not change the
 variance of the flight time, which is our primary concern.
 
 However, this quantity is only finite if the field operators have been averaged with a  test function. In the present
 context, the density profile of the slab of dielectric naturally defines a suitable function. Let $F(x)$ be a profile
 function satisfying
\begin{align}
\frac{1}{d}\int_{-\infty}^{\infty}F(x)dx = 1.
\label{norm}
\end{align}
Now the time delay operator may be written as
\begin{align}
t_d =  n_p \, \int_{-\infty}^\infty\left[1+\gamma_i\, E_i^0(\textbf{x},t)+
\mu_{ij}\, :E^0_i(\textbf{x},t)E^0_j(\textbf{x},t):\right]\,F(x)dx \,.
\label{tdF}
\end{align}
The flight time variance now becomes
\begin{align}
(\Delta t_d)^2 &= {n_p}^2\int_{-\infty}^{\infty}dx \,F(x)\int_{-\infty}^{\infty}dx' \,F(x')\Big[\gamma_i\gamma_j\langle E_i^0(\textbf{x},t)E_j^0(\textbf{x}',t')\rangle\nonumber\\
&+\mu_{ij}\mu_{lm}\langle:E_i^0(\textbf{x},t)E_j^0(\textbf{x},t)::E_l^0(\textbf{x}',t')E_m^0(\textbf{x}',t'):\rangle\Big].
\label{dtdF}
\end{align}

The definition of normal ordering, 
$$:E_i^0(\textbf{x},t)E_j^0(\textbf{x}',t'): \; = E_i^0(\textbf{x},t)E_j^0(\textbf{x}',t') - \langle E_i^0(\textbf{x},t)E_j^0(\textbf{x}',t')\rangle,$$ 
and the use of Wick's theorem lead to
\begin{align}
\langle:E_i^0(\textbf{x},t)E_j^0(\textbf{x},t): \; :E_l^0(\textbf{x}',t')E_m^0(\textbf{x}',t'):\rangle &= \langle E_i^0(\textbf{x},t)E_l^0(\textbf{x}',t')\rangle\langle E_j^0(\textbf{x},t)E_m^0(\textbf{x}',t')\rangle\nonumber\\&+\langle E_i^0(\textbf{x},t)E_m^0(\textbf{x}',t')\rangle\langle E_j^0(\textbf{x},t)E_l^0(\textbf{x}',t')\rangle.
\end{align}
Thus the flight time variance can be expressed as an integral involving the correlation functions of the electric field. 
This double integral is over the spacetime volume of the worldtube of the probe pulse wavepacket. This worldtube 
is centered upon the worldline of the middle of the wavepacket, described by $x = t/n_p$.
 We will assume that the wavepacket is sufficiently localized around this worldline so that integrations over the 
 spatial directions transverse to the $x$-direction may be neglected. 
 In this limit, we are averaging the electric field and the squared electric field along the worldline of an observer
 comoving with the probe pulse. In the rest frame of this observer, the field operators are being averaged in time
 alone. 

Once we take the coincidence limit in the transverse spatial directions, the needed electric field correlation functions for
a nondispersive, isotropic material become~\cite{BDFS15}
 \begin{align}
\langle E^0_x(x,t)E^0_x(x',t')\rangle &= \frac{1}{\pi^2\, n_b^3\, \left[(\Delta x)^2 - (\Delta t)^2/n_b^2\right]^2}\,,
\label{eq:Exx}
\\
\langle E^0_y(x,t)E^0_y(x',t')\rangle&= \langle E^0_z(x,t)E^0_z(x',t')\rangle = 
\frac{(\Delta x)^2 + (\Delta t)^2/n_b^2}{\pi^2 \, n_b^3\, \left[ (\Delta t)^2/n_b^2 - (\Delta x)^2\right]^3}\,,
\label{eq:Eyy}
\\
\langle E_i^0({x},t)E_j^0({x}',t')\rangle &= 0,\ \  i\neq j,
\end{align}
where $\Delta x = x-x'$ and $\Delta t = t -t' -i \varepsilon$, with $\varepsilon > 0$, and $n_b$ is the refractive index measured by the background field $E^0_i$.

\subsection{Fractional variance in the flight time}
Using the above results and recalling that the integrations in Eq. (\ref{dtdF}) are performed along the path of the probe pulse, 
given by $t=n_p x$, we obtain
\begin{align}
(\Delta t_d)^2 =	\int_{-\infty}^{\infty}dx \,F(x)\int_{-\infty}^{\infty}dx'F(x')\left[\frac{\alpha_1}{(\Delta x)^4}+\frac{\alpha_2 }{(\Delta x)^8}\right],
\label{final2}
\end{align}
where $\Delta x$ is understood to have a small negative imaginary part.
Here we have defined the parameters $\alpha_1$ and $\alpha_2$ as
\begin{align}
\alpha_1 &= \frac{n_b{n_p}^2}{\pi^2\left({n_p}^2-{n_b}^2\right)^2}\left[\gamma_x^2 + \left(\gamma_y^2+\gamma_z^2\right)\frac{\left({n_p}^2+{n_b}^2\right)}{\left({n_p}^2-{n_b}^2\right)}\right],
\\
\alpha_2 &= \frac{2{n_b}^2{n_p}^2}{\pi^4\left({n_p}^2-{n_b}^2\right)^4}\left[\mu_{xx}^2 + \left(\mu_{yy}^2+\mu_{zz}^2+2\mu_{zy}^2\right)\frac{\left({n_p}^2+{n_b}^2\right)^2}{\left({n_p}^2-{n_b}^2\right)^2}
+2\left(\mu_{xy}^2+\mu_{xz}^2\right)\frac{\left({n_p}^2+{n_b}^2\right)}{\left({n_p}^2-{n_b}^2\right)}\right].
\label{an}
\end{align}

This result  generalizes previous work~\cite{BDFS15}  by  including the contribution from the third order 
nonlinear susceptibility, 
and by giving an expression for the flight time for a general profile function $F(x)$.

\section{A Choice for the Switching Function}
\label{sec:switch}

We wish to choose a suitable smooth switching function that represents  the transitions which occur as the probe
pulse enters and exits  the medium. 
It will be useful to have two parameters, one ($d$) which describes the width of the slab and another ($b<d$) which describes
the effective length over which the nonlinearity changes smoothly as the pulse enters and exits.
There are several  choices for such a function. Here we use a function $F_{b,d}$ defined by
\begin{align}
F_{b,d}(x) =  \frac{1}{\pi}\left[\arctan\left(\frac{x}{b}\right)+\arctan\left(\frac{d-x}{b}\right)\right].
\label{fbd}
\end{align}
\begin{figure}[htbp]
	\centering
		\includegraphics[scale=1]{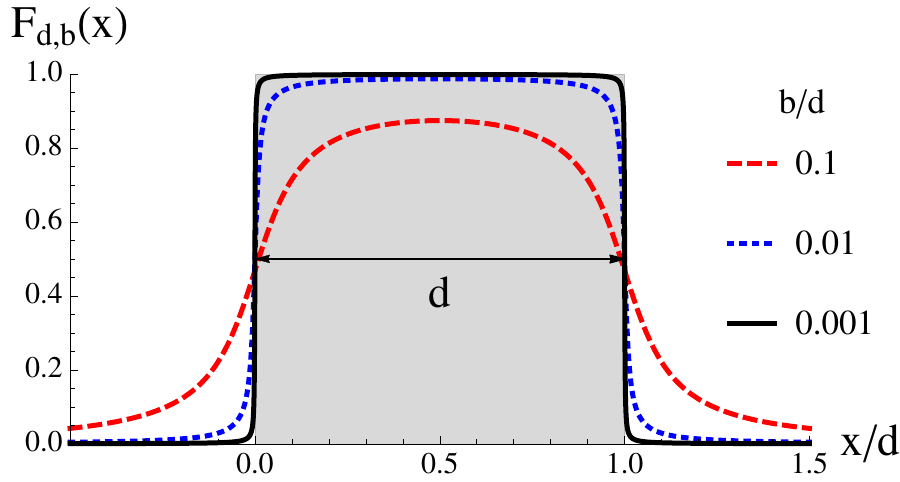}
		\caption{The switching function $F_{b,d}(x)$.}
	\label{figa}
\end{figure}	
The derivative of this function with respect to $x$ is a sum of two Lorentzian functions.
Figure \ref{figa} presents some plots of $F_{b,d}$ for a few values of the ratio $b/d$.
The parameter $b$ 
describes the distance over which $F_{b,d}(x)$ changes from its minimum to its maximum values,
and vice versa. Note that  when $b\rightarrow 0$ we recover a step function, as expected. 

The integrals appearing in Eq. (\ref{final2}) may be evaluated by contour integration, with $\Delta x = x-x' -i \varepsilon$, 
where $\varepsilon > 0$. The results, and their asymptotic forms when  $b \ll d$, are 
\begin{align}
&\int_{-\infty}^{\infty}\int_{-\infty}^{\infty}dxdx'F_{d,b}(x)F_{d,b}(x')\frac{1}{\Delta x^4} = 
\frac{d^2(d^2+12b^2)}{12b^2(d^2+4b^2)^2} \sim \frac{1}{12 b^2}  \, ,\\
&\int_{-\infty}^{\infty}\int_{-\infty}^{\infty}dxdx'F_{d,b}(x)F_{d,b}(x')\frac{1}{\Delta x^8} = \frac{d^2(21504b^{10}+1344b^6d^4+240b^4d^6+24b^2d^8+d^{10})}{1344b^6(4b^2+d^2)^6} \sim \frac{1}{1344 b^6}  \,  .
\label{eq:b6}
\end{align}   
 If we assume $b \ll d$, and use the above asymptotoic forms, we obtain
\begin{align}
(\Delta t_d)^2 \approx \frac{\alpha_1}{12b^2}+\frac{\alpha_2 }{1344b^6}.
\label{final3}
\end{align}
We define the squared fractional variance in flight time of the probe field as
\begin{align}
\delta^2 = \frac{(\Delta t_d)^2 }{\langle t_d\rangle^2} 
\approx \frac{\alpha_1}{12{n_p}^2d^2b^2}+\frac{\alpha_2 }{1344{n_p}^2d^2b^6}.
\label{frac}
\end{align}

\begin{figure}[htbp]
	\centering
		\includegraphics[scale=1]{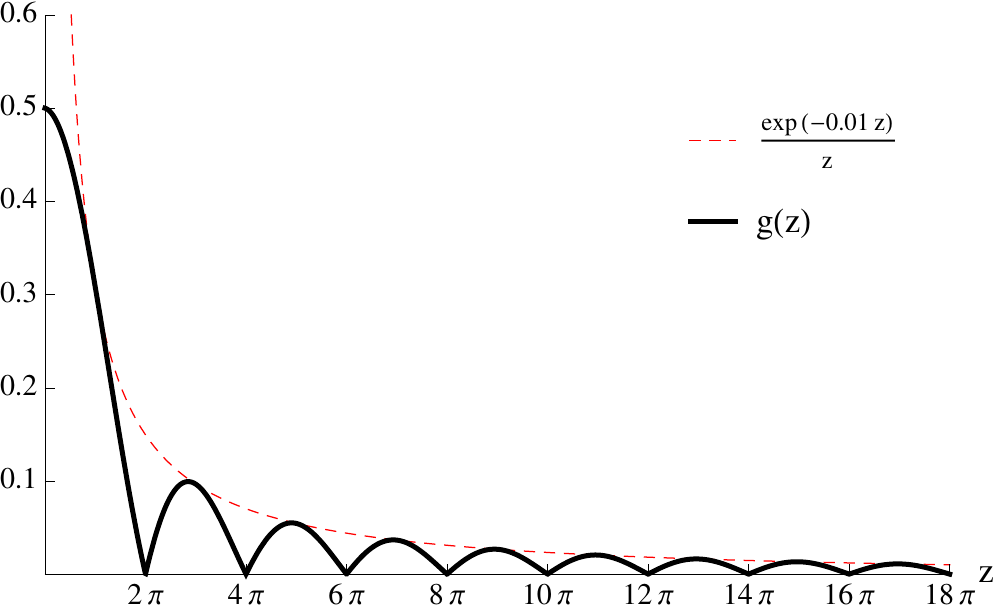}
		\caption{Modulus of the Fourier transform of  $F_{b,d}(x)$ is illustrated. Specifically, the
		function $g(z)=\sqrt{\pi/2}\,|\hat{F}_{b,d}(k)|/d$ is plotted as a function of $z=kd$ for the case $b=0.01$}
	\label{figb}
\end{figure}

The modulus of the Fourier transform of $F_{b,d}$ is given by 
\begin{align}
\left| \hat{F}_{b,d(k)}\right| = \frac{1}{k}\sqrt{\frac{2}{\pi}}\left|\sin\frac{kd}{2}\right|\, e^{-|k|b},
\label{fourier}
\end{align}
and its behavior is depicted in Fig. \ref{figb}, where we defined the dimensionless variable $z=kd$ and 
function $g(z)=\sqrt{\pi/2}\,|\hat{F}_{b,d}(k)|/d$.  
Note that $\lim_{z\rightarrow 0} g(z) = 1/2$ and that $g(z)$ falls exponentially as $z$ increases. The plot was 
done with the particular choice $b=0.01d$, for which more than $90\%$ of total area under the solid curve occurs in the 
range $0\le z \le 18\pi$.

Recall that our approximations require (i) Eq.~(\ref{eq:C1}), the dominance of the vacuum field over the probe
field, (ii)  Eq.~(\ref{eq:C2}), which is equivalent to $\lambda_p < \lambda_b$, 
(iii) a range of frequencies in which the material can  be assumed free of dispersion, and (iv) a material which is 
approximately isotropic, at least for the frequencies which give 
the primary contribution to the background field. The rate of decay of the Fourier transform $\hat{F}_{b,d}(k)$
allows us to test approximations (ii) and (iii).
 The exponentially decreasing behavior of the Fourier transform of this function, depicted in Fig.~\ref{figb},
 suppresses the high energy modes of the background field.

For the case $b=0.01d$, at least $90\%$  of the effect will occur in the range $0 \le z \lesssim 18\pi$, which means 
that only wave lengths such that $\lambda_b \gtrsim d/9$ will significantly contribute. For a slab with 
$d \approx 10\mu m$, the dominant wavelengths of the background field are those with 
$\lambda_b \gtrsim 1.1\mu m$. Shorter wavelength modes are naturally suppressed by the time averaging.
Furthermore,  the larger contribution  occurs arises from  $z \leq2\pi$, which for $b=0.01d$, corresponds to 
a wavelength of $\lambda_b \approx 10\mu m$.  Thus if the material is relatively free of dispersion when
$\lambda_b \gtrsim 1.1\mu m$, then our assumption that $n_b$ is independent of frequency is justified.
We may choose $\lambda_p  \alt 1 \mu m$ to satisfy Eq.~(\ref{eq:C2}).

We may also justify the assumption of the dominance of the vacuum field over the probe
field, Eq.~(\ref{eq:C1}), using essentially the same argument as was given in Sec.~3.2 of Ref.~\cite{BDFS15}.
The only difference is that in Ref.~\cite{BDFS15}, $E_0^2 \propto 1/\tau^4 \propto 1/d^4$ is the expectation
value of the square of the averaged electric field. Here it is the square root of the expectation value of the square 
of the averaged squared electric field, which can be obtained from Eq.~(\ref{eq:b6}) divided by $d^2$, and
is proportional to $1/(b^3\,d)$ for $b \alt d$. 
Thus, if $b\approx d$, the two quantities are of the same order, and we obtain Eq.~(30) of Ref.~\cite{BDFS15} as 
the condition that the vacuum field dominate the probe field. If $b < d$, then the vacuum field is enhanced
by the shorter switch-on and switch-off times,
and it becomes easier to satisfy Eq.~(\ref{eq:C1}). The physical reason for vacuum dominance is that many
more modes contribute to the vacuum field than to the probe field.

\section{Estimates}
\label{sec:estimates}
The first example we wish to study is the crystal of cadmium selenide (CdSe), which is a hexagonal one, point 
group 6mm. This system was already investigated \cite{BDFS15} in the  case of a Lorentzian
sampling  function.  CdSe is an optical medium with nonzero 
 second order nonlinear dielectric susceptibilities and satisfies the conditions  discussed at the end of last section. 
This crystal has an index of refraction $n_b = 2.43$ and  a second order coefficient $\chi^{(2)}_{zzz} \approx 1.1\times 10^{-10}mV^{-1}$ at a wavelength $\lambda_b = 10.6 \mu m$ \cite{patel1966,Charra2000}. Now setting the wavelength of the probe field as $\lambda_p = 1.06 \mu m$, for which $n_p = 2.54$, and setting the parameter $b = 0.01 d$, we obtain from Eq. (\ref{final3}) a fractional variance of the flight time,
\begin{align}
\delta \approx 1.3\times 10^{-6} \left(\frac{10\mu m}{d}\right)^2.
\label{CdSe}
\end{align}
Compared to the model where an idealized Lorentzian distribution \cite{BDFS15} is used, this result shows
 that in the situation described by $F_{b,d}$, with $b=0.01 d$, the predicted effect is about 100 times stronger. 
  This enhancement arises because the contribution to $\delta$ due to linear electric field fluctuations is 
  proportional  to $1/(b\, d)$, as may be seen from the first term on the right-hand-side of Eq.~(\ref{frac}). 

Now we investigate a third order nonlinear optical material. Silicon (Si) is a centrosymmetric crystal (point group m3m), which means that the second order nonlinear dielectric susceptibilities are identically zero.
This crystal has  a third order coefficient $\chi^{(3)}_{zzzz} \approx 2.80\times 10^{-19}m^2V^{-2}$ at a wavelength 
$\lambda_b = 11.8 \mu m$~\cite{wynne1968,Charra2}, and an index of refraction $n_b = 3.418$ at the same 
wavelength~\cite{Salzberg57}.
 Suppose the probe wave packet has a peak wavelength of 
$\lambda_p = 1.4 \mu m$, for which $n_p = 3.484$~\cite{Primak71}. As before, using $b = 0.01 d$,  we find the dominant 
contribution to the fractional variance of the flight time,
\begin{align}
\delta \approx 4.2 \times 10^{-8} \left(\frac{10\mu m}{d}\right)^4.
\label{Si}
\end{align}
Note that the contribution to $\delta$ due to quadratic electric field fluctuations is 
  proportional  to $1/(b^3\, d)$, as may be seen from the second term on the right-hand-side of Eq.~(\ref{frac}). 
As expected, the effect produced by the third order coefficient tends to be smaller than that related to second order nonlinearities. 
It may be possible to increase the effects of quadratic fluctuations if new materials with larger third order susceptibilities
can be found. In the next section, we will discuss a different type of enhancement.

\section{Probability of Large Fluctuations}
\label{sec:prob}

In the previous sections, we have been concerned with the variance of the flight time, which is in turn determined
by the variance of the  sampled electric field or of the squared electric field. Here we wish to estimate the
probability of much larger fluctuations than those described by the variance. In the case of effects produced
by the second order polarizability, this probability will be very small, as the probability distribution for fluctuations
of the electric field is Gaussian, and hence falls very rapidly. However, flight time variations due to the third
order polarizability will be associated with a more slowly decreasing probability distribution. The distributions
for quadratic quantum operators have been discussed in Refs.~\cite{FFR10,FFR12,FF15}. In particular,
the asymptotic form for the probability distribution of the Lorentzian average of the squared electric field was
given in Ref.~\cite{FFR12}. The sampling function used in this paper,  Eq.~(\ref{fbd}), is not Lorentzian, but the
magnitude of its Fourier transform,  Eq.~(\ref{fourier}), has the same exponential decay as in the Lorentzian
case. Furthermore, it was argued in  Ref.~\cite{FF15} that the decay rate of the  Fourier transform of the sampling
function determines the asymptotic form for the probability distribution. Thus it is reasonable to extrapolate 
the Lorentzian results to the present case.

Here we briefly summarize the needed results from  Ref.~\cite{FFR12}.  Let $:\bar{E^2}:$ be the Lorentzian 
time average of the normal ordered squared electric field operator at a given point in space, or more generally
along a timelike worldline. In our problem, this will be the path of the probe wave packet. Define the 
dimensionless variable
\begin{equation}
x = (4 \pi \, \tau^2)^2 \, :\bar{E^2}: \,,
\end{equation}
 where $\tau$ is the characteristic averaging time. Let $P(x)$ be the probability distribution for finding a given
 value of $x$ in a measurement in the vacuum state, which is normalized by
\begin{equation}
\int_{-x_0}^\infty P(x)\, dx =1\,.
\end{equation}
Here $-x_0$ is the lower bound, the smallest value of $x$ which could ever be observed. Note that this lower
bound is negative, so measurements of $ :\bar{E^2}: $ in the vacuum state can return negative values, just
as expectation values of the squared electric field in more general states can be negative. In fact, one expects
most measurements in the vacuum to result in a negative value, but when the outcome is positive, it is likely to
be larger in magnitude. Note that a negative value of  $:\bar{E^2}:$ results in a time {\it advance} compared
to the mean flight time in the material, just as positive values result in time delays.  

Our primary interest is in
the asymptotic form of $P(x)$ when $x \gg 1$, which describes the probability of finding especially large
values of the  squared electric field. This asymptotic form is approximately
\begin{equation}
P(x) \sim c_0 x^{-2}\, {\rm e}^{-a\, x^{1/3}} \,,
\label{eq:P}
\end{equation}
where $c_0 \approx 0.955$ and $a \approx 0.764$. A striking feature of this result is the one-third power
in the exponential, which causes $P(x)$ to fall much more slowly than a Gaussian or an exponential function.
Given $P(x)$, we can define the cumulative probability distribution by
\begin{equation}
{\cal P}(y) = \int_y^\infty P(x)\, dx \,,
\end{equation}
which gives the probability of finding any value greater than or equal to $y$ in a given measurement.
If $y \gg 1$, we can directly integrate Eq.~(\ref{eq:P}) to find
\begin{equation}
{\cal P}(y) \approx \frac{3 c_0}{a y^{4/3}} \; {\rm e}^{-a\, y^{1/3}} \,.
\label{eq:Pcumm}
\end{equation}

It is shown in Ref.~\cite{FFR12} that the second moment of $P(x)$ for the squared electric field is 
\begin{equation}
\mu_2 = \int_{-x_0}^\infty x^2\, P(x)\, dx = 6\, ,
\end{equation}
so the root mean square of $x$ is $x_{rms} = \sqrt{6}$. Now we may use Eq.~(\ref{eq:Pcumm}) to find the
probability of a result which exceeds a large multiple of  $x_{rms}$. Some examples are
given in Table~\ref{table:probs}.

\begin{table}[htbp]
\caption{Probabilities of large squared electric field fluctuations. }
\label{table:probs}
\begin{center}
\begin{tabular}{|c|c|c|c|c|c|} \hline
$y$  & ${\cal P}(y)$   \\ \hline 
 $10 \, x_{rms}$ & $0.006$ \\ \hline
 $100 \, x_{rms}$ & $2.1 \times 10^{-5}$ \\ \hline
 $10^3 \, x_{rms}$ & $4.0 \times 10^{-9}$ \\ \hline
  $10^4 \, x_{rms}$ & $1.3 \times 10^{-15}$ \\ \hline
\end{tabular}
\end{center}
\end{table}

The same probabilities apply to the flight time delay due to vacuum squared electric field fluctuations.
Thus there is a probability of about $4.0 \times 10^{-9}$ that a given pulse will suffer a delay which is
1000 times larger than the root mean square value, given for example by Eq.~(\ref{Si}). Note that our
discussion is rather heuristic, and these are order of magnitude estimates. In particular, we have not made
a clear distinction between the squared electric field in the rest frame of the probe pulse, and that in
the rest frame of the dielectric material. However, for $n_p \approx 3$, so $v_{ph} \approx 1/3$ to leading
order, these quantities will be of the same order. 

Another important point is that both the Lorentzian function and the function $F_{b,d}$ defined in
Eq.~(\ref{fbd}) have tails in both directions. A more realistic choice is a function of compact support,
which is strictly zero before the measurement process begins. Such functions lead to even slower
decrease of the probability distribution for large arguments~\cite{FF15}.

\section{Summary and Discussion}
\label{sec:sum}

In this paper, we have extended previous work~\cite{FDMS13,BDFS15} on analog models for lightcone 
fluctuations. Here we have been concerned with vacuum fluctuations of the quantized electric field,
and especially with fluctuations of the squared electric field. Our model studies the flight time of a
probe pulse through a material with a non-zero third order polarizability. Vacuum fluctuations of the
squared electric field lead to fractional flight time variation which can be of order $ 4 \times 10^{-8}$ in the
example given in Sec.~\ref{sec:estimates}.  These flight time variations model the effects of the passive
spacetime geometry fluctuations driven by quantum stress tensor fluctuations.

We have also extended the study of the effects of temporal switching functions on the  fluctuations of 
quadratic quantum operators. In Sec.~\ref{sec:switch}, we discussed a specific choice of switching function,
which can be relatively constant over a finite interval, and can model the density profile of the nonlinear
material. The Fourier transform of this function falls exponentially at a rate determined by the parameter
$b$, which controls the rate of rise and fall at the ends of the plateau of this function. As this parameter
is decreased, increasingly higher frequency modes contribute and increase the fractional flight time variation.
We were able to use
the Fourier transform to estimate the range of vacuum modes which contribute to the flight time variation
and to test our approximation of ignoring dispersion. 

In Sec.~\ref{sec:prob}, we discussed the probability of especially large fluctuations in flight time. This
analog model may provide a means to study the probability of large stress tensor fluctuations, which
tend to fall more slowly than a Gaussian function~\cite{FFR10,FFR12,FF15}. We estimated, for example, 
a probability of $2\times 10^{-5}$ for finding a flight time delay which is at least $100$ times the typical delay.

There are then two distinct signatures of squared electric field fluctuations. The first is the fractional flight time 
variation $\delta$ estimated, for example, in Eq.~(\ref{Si}). The second is the pulses which undergo an
especially long time delay due to a very large  squared electric field fluctuation. The extent to which either
can be observed in a realistic experiment is a topic for future work. One aspect of this work will be an
exploration of finite duration switching, which is more realistic than functions with tails extending into the
past and the future. Such functions with compact support were treated in Ref.~\cite{FF15}, where it
was shown that quadratic quantum operators averaged in time with such functions are associated
with a probability distribution which falls more slowly than that for the Lorentzian, Eq.~(\ref{eq:P}).
This raises the possibility that large fluctuations can be more likely than was estimated in Sec.~\ref{sec:prob}.

\begin{acknowledgments}
This work was supported in part by the National Science Foundation under grant PHY-1506066,
and by the Brazilian research agencies CNPq (grants 304486/2012-4 and 168274/2014-0), FAPEMIG (grant ETC-00118-15),
 and CAPES.
\end{acknowledgments}

\end{document}